\pdfoutput=1

\documentclass[aps,prl,twocolumn,groupedaddress,amsmath,amssymb]{revtex4-1}

\usepackage{graphicx}
\usepackage{color}
\usepackage{amsmath}
\usepackage{gensymb}

\begin{document}

%\preprint{}

\title{Shear Destruction of Frictional Aging and Memory}
%partial slip mediates frictional aging


\author{Sam Dillavou$^1$, Shmuel M. Rubinstein$^2$}
%\thanks{}
\affiliation{$^1$Physics, Harvard University, Cambridge, Massachusetts 02138, USA \\
$^2$Applied Physics, Harvard University, Cambridge, Massachusetts 02138, USA}

%used with the \author comman(d).

\date{\today}

\begin{abstract}
We simultaneously measure the static friction and the real area of contact between two solid bodies. Under static conditions both quantities increase logarithmically in time, a phenomenon coined aging. Indeed, frictional strength is traditionally considered equivalent to the real area of contact. Here we show that this equivalence breaks down when a static shear load is applied during aging. The addition of such a shear load accelerates frictional aging while the aging rate of the real area of contact is unaffected. Moreover, a negative static shear - pulling instead of pushing - slows frictional aging, but similarly does not affect the aging of contacts. The origin of this shear effect on aging is geometrical. When shear load is increased, minute relative tilts between the two blocks prematurely erase interfacial memory prior to sliding, negating the effect of aging. Modifying the loading point of the interface eliminates these tilts and as a result frictional aging rate becomes insensitive to shear. We also identify a secondary memory-erasure effect that remains even when all tilts are eliminated and show that this effect can be leveraged to accelerate aging by cycling between two static shear loads.
\end{abstract}

%\keywords{}

\maketitle

%\section{Section 1}
%\subsection{Subsection 1}
%\subsubsection{Subsubsection1}

\newcommand{\appropto}{\mathrel{\vcenter{
  \offinterlineskip\halign{\hfil$##$\cr
    \propto\cr\noalign{\kern2pt}\sim\cr\noalign{\kern-2pt}}}}}

The static coefficient of friction, $\mu_S$, is traditionally considered a property of the materials that compose an interface. However, a wide range of experiments show that $\mu_S$ depends on far more than the material(s) involved, including environmental conditions \cite{Berthoud:1999ha, Bocquet:1998wt,Frye:2002jja}, loading rate \cite{Karner:1998ik,Karner:2000aa, Karner:2001cg}, and geometry \cite{Yamaguchi:2016ga, BenDavid:2011di}. Because even macroscopically flat materials are typically rough at small scales, frictional interfaces are composed of myriad discrete microcontacts \cite{BowdenTabor}. The instantaneous frictional resistance is therefore a function of the state of this multicontact interface (MCI), a relationship captured by the Rate and State Laws \cite{Dieterich:1979vq, Rice:1983aa, Ruina:1983hh}. For example, under set conditions the state of the interface evolves in time such that in most cases, the longer two bodies are in contact, the harder it is to induce sliding \cite{Rabinowicz:bUxLFTVs, Dieterich:1972ta}. While extremely useful in many circumstances, these laws are phenomenological and thus do not describe the physical underpinning of frictional resistance.

For a wide range of materials, from rock \cite{Dieterich:1972ta} to paper \cite{Heslot:1994gd} to metal \cite{Rabinowicz:bUxLFTVs} to plastic \cite{Berthoud:1999ha} and even granular materials \cite{Karner:1998ik}, the static coefficient of friction grows logarithmically in time under constant normal load, often referred to as aging. This evolution is most often attributed to an increase in the real area of contact, $A_R$, within the interface \cite{Dieterich:1994ux, Baumberger:2006bq}, however other mechanisms such as chemical bonding \cite{Li:2011gf}, or formation of capillary bridges \cite{Bocquet:1998wt} have been shown to produce similar effects. Recent work \cite{Dillavou:2018in} has shown that this logarithmic behavior is in fact real aging; the interface exhibits memory, and behaves phenomenologically similarly to a large class of glassy, disordered systems \cite{Amir:2012fe, Lahini:2017dq, Dillavou:2018in}. Notably, this means that the history of an interface, locally and globally, may be gleaned from its subsequent evolution in time \cite{Dillavou:2018in}. 

In many frictional interfaces, the growth rate of $\mu_S$ is increased by the presence of a static shear load $S_0$  \cite{MasaoNakatani:1996wu,Berthoud:1999ha,Bureau:2002br}, the cause of which is still debated. The simplest possibility is that static shear accelerates the evolution of contact patches, although empirical evidence for this effect is inconclusive \cite{Ryan:2018kc}. An alternative hypothesizes that raising shear to measure $\mu_S$ modifies the contact area. This modification has been speculated to result from micro-slips consisting of individual contacts detaching \cite{Berthoud:1999ha}, and from deformation of the contacts modifying their area \cite{Nagata:2012hv}. After aging, any such alteration of contact will weaken the overall interfacial strength; old contacts that have expanded over time are replaced with new, un-aged, and thus smaller, contacts. Both scenarios, are consistent with the current paradigm, but neither has experimentally confirmed. It is therefore still unclear why a constant shear load accelerates the aging of an interface, although resolving the origin of this effect may have significant implications for the physics behind friction.

Here we experimentally demonstrate that shear-accelerated aging is in fact a consequence of minuscule tilts between samples. These tilts are a consequence of small torque imbalances, and result in  redistribute normal pressure, destroying aged contacts and creating fresh ones. That is, it is the change in shear, not its static value, that influences the evolution of an interface. We show that the growth rate of $\mu_S$ is linearly dependent on static shear. This effect extends into negative shear (pulling), ruling out the hypothesis that shear can only accelerate aging. Furthermore we measure that the evolution of the total area of contact, $A_R$, is insensitive to static shear load. By minimizing the tilts that redistribute contact we eliminate shear-accelerated aging, confirming that the effect is geometrical. Nevertheless, we show that even without tilts, a change in shear erases interfacial memory through a secondary effect.

\begin{figure} %%%%%%%%%%%%%%%%%%%%%%%%%%%%%%%%%%%%%%
\hspace*{-.15in}
\includegraphics[width = .5\textwidth]{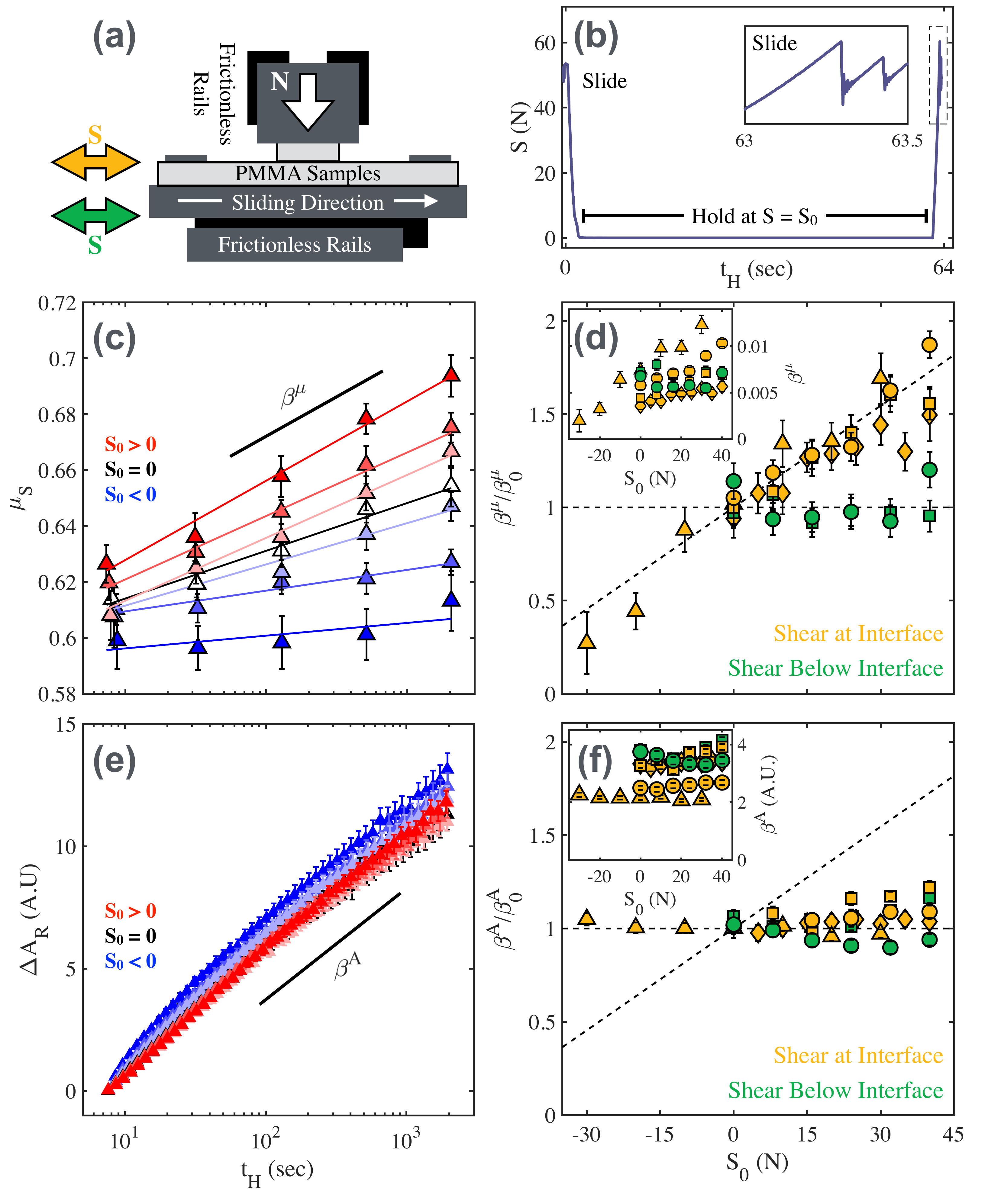}
\caption{Slide-hold-slide experiments. (a) Schematic of the biaxial compression/translation stage. Experiments are conducted using one of two shear application points, as shown by the horizontal arrows. (b) Typical experimental slide-hold-slide (SHS) protocol for measuring static friction. $A_R$ is measured during the hold period, as indicated. Inset: typical example of static frictional peak. (c) $\mu_S$ versus $t_H$ for $S_0 = \{-30, -20, -10, 0, 10, 20, 30\}N$. Color represents $S_0$ value. In this experiment, shear is applied at the level of the interface (yellow arrow in (a).) (d) $\beta^\mu(S_0)/\beta^\mu_0$ versus $S_0$. Colors correspond to shear application points in (a), shapes correspond to sample pairs, and both are consistent in Fig \ref{1} and \ref{2}. Dashed lines correspond to $\epsilon^\mu \sim 0.017 $N$^{-1}$ and $0 $N$^{-1}$, and are repeated in (f). Inset: $\beta^\mu$ versus $S_0$. (e) Area of contact $A_R$ in arbitrary units \cite{Dillavou:2018in} versus $t_H$ for the same experiment as (c). (f) $\beta^A(S_0)/\beta^A(S_0 = 0)$ as a function of $S_0$, for the same experiments as (d). Inset: $\beta^A$ as a function of $S_0$.
\label{1}}
\end{figure} %%%%%%%%%%%%%%%%%%%%%%%%%%%%%%%%%%%%%%%

We simultaneously measure the static friction coefficient and the real area of contact across an entire 2D interface. The interface is formed between two laser-cut PMMA (poly methyl-methacrylate) blocks with 1.5 - 2.5 cm$^2$ of nominal contact area. The bottom block is original, extruded PMMA, and the top block is lapped with 1000 grit polishing paper. Samples are washed with soap, rinsed with deionized water then isopropanol, and finally air dried. The samples are held by a biaxial compression/translation stage, as shown in Fig \ref{1}(a).The interface is illuminated using total internal reflection (TIR) and imaged through the top sample. These images are used to calculate the real area of contact, as described in detail in previous work  \cite{Dillavou:2018in}. Shear is applied either directly at the level of the contact plane, or 3cm below, as indicated in Fig \ref{1}(a). All experiments are conducted at a constant normal load, $N = 90$N, and samples are slid prior to each experiment at 0.33 mm/s to reset the interface. To bypass any systematic wear effects, a randomization and averaging protocol is implemented for the measurements of the coefficient of static friction, identical to the protocol described in \cite{Dillavou:2018in}.

The gradual increase in the static friction coefficient over time has been demonstrated experimentally for a plethora of materials and systems \cite{Dieterich:1972ta,Heslot:1994gd,Rabinowicz:bUxLFTVs,Berthoud:1999ha,Karner:1998ik}. The standard experimental test of this effect is a Slide-Hold-Slide (SHS) protocol under constant normal load, $N$. In this procedure an interface is slid, then held at a constant shear $S_0$ for time $t_H$, then slid again to measure the static friction coefficient $\mu_S$, as shown in Fig \ref{1}(b). For a given $S_0$, $\mu_S$ grows logarithmically in time as
\begin{equation}
\mu_S = \mu_0 + \beta^\mu \log(t_H)
\label{eq1}
\end{equation} 
as shown in Fig \ref{1}(c), where $\mu_0$ is a reference static friction value and $\beta^\mu$ is the frictional aging rate. $\beta^\mu$ was previously reported to increase under a shear load \cite{MasaoNakatani:1996wu,Berthoud:1999ha,Bureau:2002br}, however these studies only examined positive values of $S_0$, i.e. pushing. Surprisingly, we find that negative values of $S_0$, i.e. pulling, decrease $\beta^\mu$ compared to the shear-free case, as shown in Fig \ref{1}(c). In fact, $\beta^\mu$ is linear in $S_0$ and is described by
\begin{equation}
    \beta^\mu = \beta^\mu_0 \big(1 + \epsilon^\mu S_0\big) 
\end{equation}
as shown in Fig \ref{1}(d), where $\beta^\mu_0$ is the frictional aging rate at $S_0 = 0$, and $\epsilon^\mu$ is a constant. While $\beta^\mu_0$ varies significantly, $\epsilon^\mu$ is approximately constant across all samples. However, $\epsilon^\mu$ is not only a material property but a geometrical one; lowering the shear application point by 3cm eliminates the effect of shear completely, reducing $\epsilon^\mu \rightarrow 0$, as shown in Fig \ref{1}(d). Applying shear at this level approximately balances torque-induced tilting of the interface, as we discuss shortly.

\begin{figure} %%%%%%%%%%%%%%%%%%%%%%%%%%%%%%%%%%%%%%
\hspace*{-.15in}
\includegraphics[width = .5\textwidth]{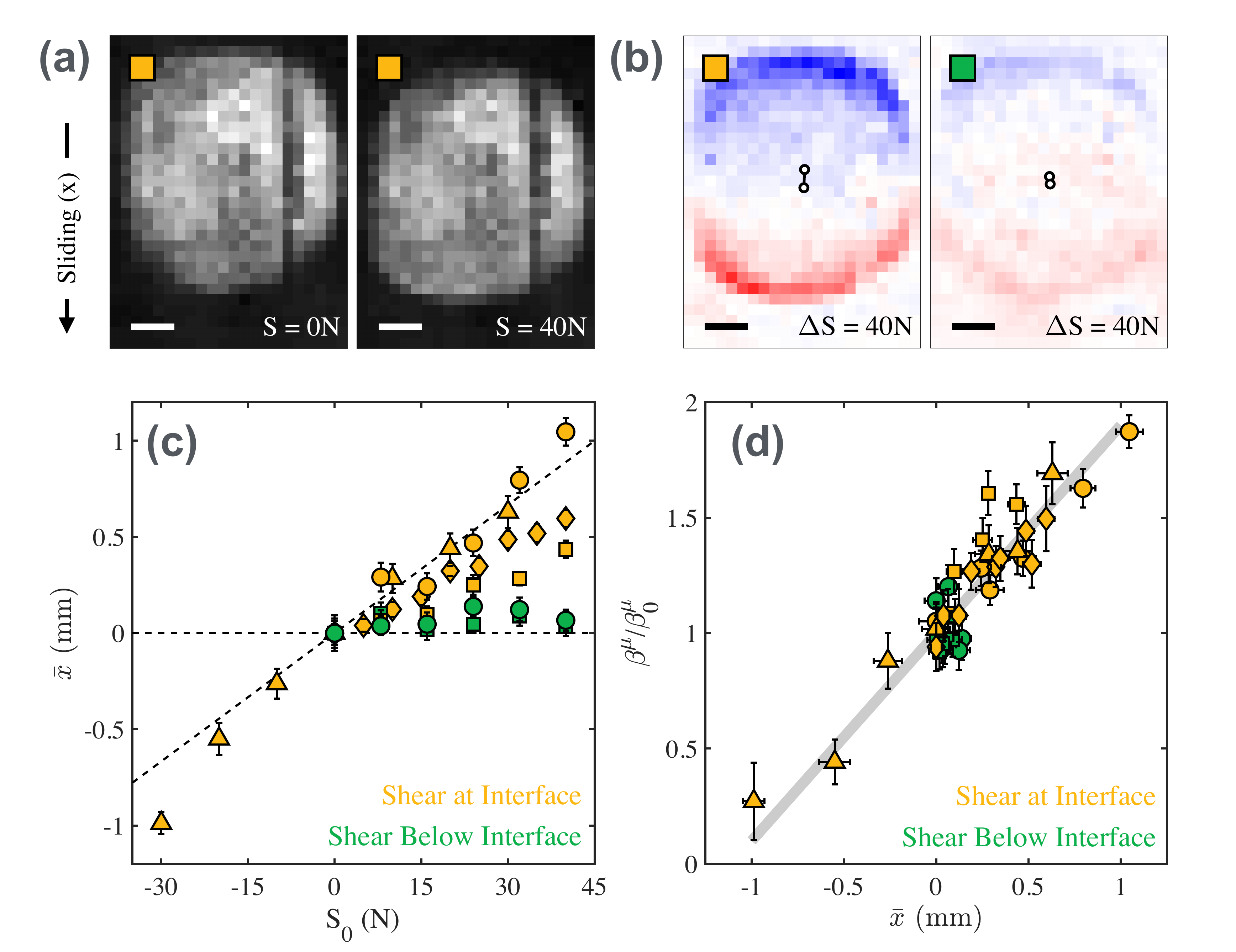}
\caption{Contact redistribution drives shear-accelerated aging. (a) Typical images of the interfacial contact plane after background subtraction, binned for visual clarity. Images are taken at static shear load $S=0$ and $S=40N$, respectively, applied at the level of the interface. Scale bars are 2mm. (b) Subtraction of interfacial images at $S=0$ and $40N$ for shear load applied at the interface (yellow) and 3cm below the interface (green). Red indicates positive change, blue negative. Center of mass, $(\overline x,\overline y)$, for the two subtracted two images are superimposed as hollow black circles. (c) $\overline x$ versus $S_0$. Shapes correspond to samples, colors to shear application position, as in Fig \ref{1} and (d). (d) $\beta^\mu/\beta^\mu_0$ versus $\overline x$ for four sample pairs and approximately 4000 total experiments.
\label{2}}
\end{figure}

Like $\mu_S$, the real area of contact, $A_R$, grows logarithmically in time as
\begin{equation}
 A_R = A_0 + \beta^A \log(t_H)
\label{eq2}
\end{equation}
as shown in Fig \ref{1}(e), where $\beta^A$ is the contact aging rate, and $A_0$ is a reference area. In contrast to $\beta^\mu$, $\beta^A$ does not depend on the imposed shear, regardless of loading position, as shown in Fig \ref{1}(e) and (f). Such a stark discrepancy between the evolution of $\mu_S$ and $A_R$ is inconsistent with the classical view of friction. This seemingly paradoxical result is resolved through geometrical considerations, taking into account the extended nature of the frictional interface; when shear is applied at the level of the interface, the top block experiences a torque and tilts. This results in a redistribution of normal load, and thus of real area of contact, as shown in Fig \ref{2}(a) and (b). This is likely a general effect in frictional sliding. Flat interfaces are extremely sensitive to angle changes; in our system tilting the bottom sample by $0.01 \degree$ replaces up to 20\% of the total area of contact. Applying shear 3cm below the interface balances the torques such that the samples tilt in tandem and the interface is undisturbed, as shown in Fig \ref{2}(b).

Without torque balance, a change in shear redistributes contact primarily along the direction of sliding, $x$. This can be quantified via the `center of contact' in $x$
\begin{equation}
    \overline x \equiv \frac{\iint xI \ dxdy}{\iint I \ dxdy} - C
\end{equation}
where $I = I(x,y)$ is the intensity of an image of interfacial contact, and $C$ is a constant guaranteeing that $\overline x(S_0=0) = 0$. $\overline x$ is linear and monotonic in $S_0$, which is similar to the frictional aging rate $\beta^\mu$, as seen by comparing Fig \ref{1}(d) and Fig \ref{2}(b). Indeed, the two values are equivalent, as shown in Fig \ref{2}(c).

 %%%%%%%%%%%%%%%%%%%%%%%%%%%%%%%%%%%%%%%

\begin{figure} %%%%%%%%%%%%%%%%%%%%%%%%%%%%%%%%%%%%%%
\hspace*{-.15in}
\includegraphics[width = .5\textwidth]{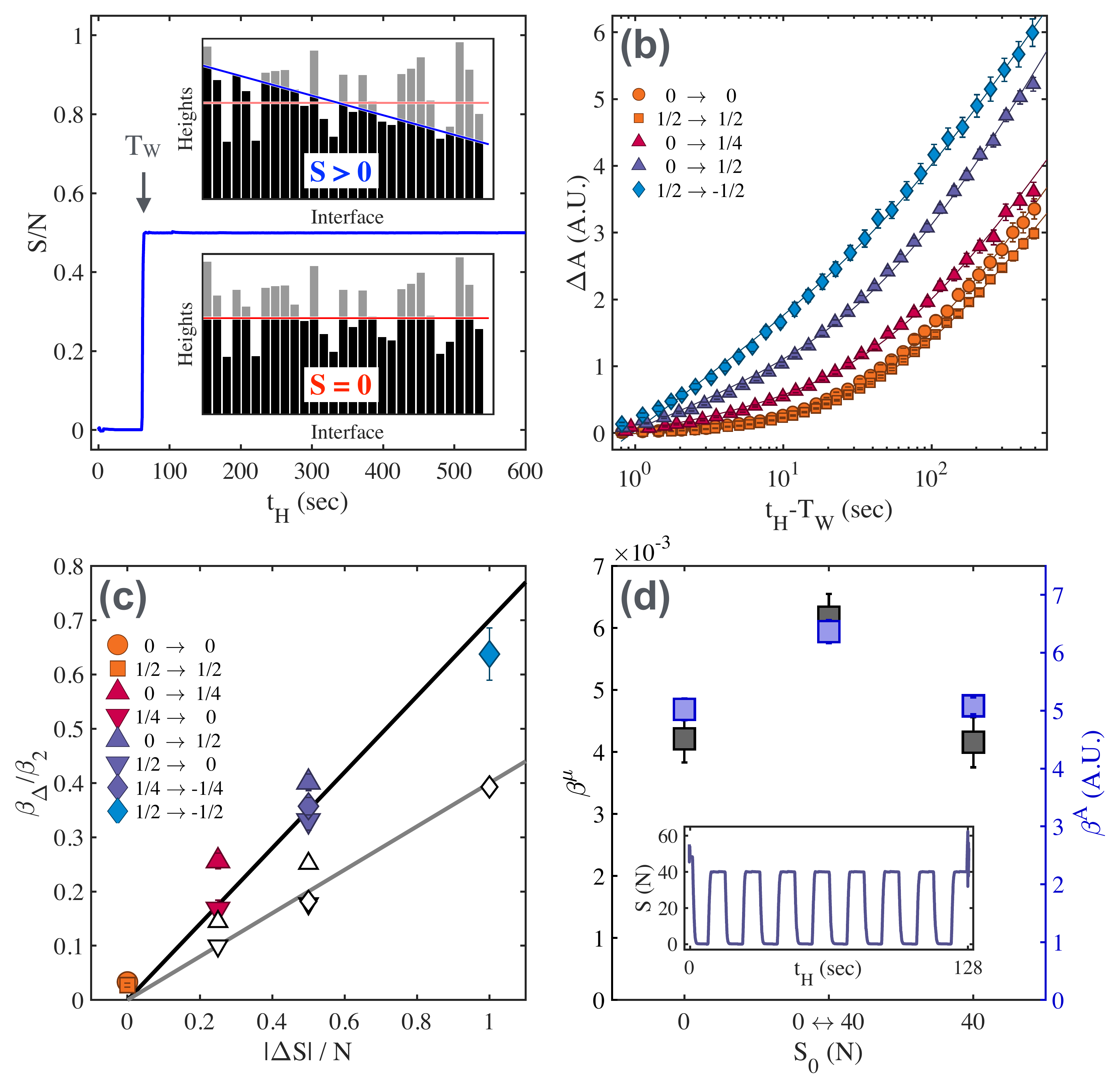}
\caption{A change in shear erases interfacial memory in two ways. (a) A typical two-step protocol in shear with constant normal load, $t_h = 60s$ and $S/N = 0$ to $S/N = 1/2$. Insets: the effect of a tilted interface is shown schematically. Note the portions of the interface on the left that detach entirely. (b) Mean values of $\Delta A \equiv A - A(T_W)$ for five distinct two-step shear protocols with constant normal load. Lines are fits to eq \ref{twoStep}. Colors and shapes are consistent in (b) and (c), and the legend indicates values of $S/N$. (c) $\beta_\Delta / \beta_2$ for eight distinct two-step protocols in shear plotted versus absolute change in shear. Black hollow symbols represent $\beta_\Delta / \beta_2$ measured only from regions of the interface with less than 5\% change in total contact when the shear load is changed. Lines are guides for the eye. (d) $\beta^\mu$ and $\beta^A$ for two standard slide-hold-slide experiments at $S_0 = 0$ and $40N$, and one experiment (in the middle) wherein $S$ was cycled between $0N$ and $40N$ every 8 seconds, as shown for a typical example in the inset.
\label{3}}
\end{figure}

The equivalence between $\beta^\mu$ and $\overline x$ is a geometrical effect, wherein different regions of the interface experience different loading history. Thus, at the time of sliding, the frictional strength is dependent on the ensemble of varied local states across the interface. We recently proposed a simple linear model for interfacial dynamics \cite{Dillavou:2018in}, which accounts for complex loading history. In this framework, the interface is composed of a heterogeneous ensemble of contacts, where the quantity of contact and rate of contact growth are proportional to the local normal stress, and entirely ignorant of shear. In a system that evolves under constant $S_0$ and $N$ for any time $t_H$, this model predicts $\beta^A \propto A_R \propto N$, regardless of contact distribution, consistent with the results presented in Fig \ref{1}(f). However, if, after this evolution, the shear is rapidly changed, the situation becomes more complex. This is exactly the case when shear is increased to the point of failure to measure $\mu_S$. Shear-induced tilts modify local normal stresses across the interface, detaching aged contacts in some regions while forming new contacts in others, as shown schematically in Fig \ref{3}(a). As a result, the interface is partially refreshed, weakening the effect of aging and reducing $\beta^\mu$, consistent with the results presented in Fig \ref{1}(d).

Sliding to measure $\mu_S$ largely resets the interface. However, if instead of increasing shear to the point of failure, it is raised to a constant high value $S_2$, as shown for a typical example in Fig \ref{3}(a), the interface retains a memory of its past states. As a result, the local evolution of the area of contact will be a consequence of its unique loading history. Our model predicts that after such a protocol, the real area of contact will evolve as
\begin{equation}
 \Delta A_R(t) = (\beta^A -\beta_\Delta) \log(t) + \beta_\Delta \log(t-t_H)
\label{twoStep}
\end{equation}
consistent with our measurements, as shown in Fig \ref{3}(b). $\beta^A$ is the familiar constant logarithmic aging rate of the entire system, and $\phi \equiv \beta_\Delta/\beta^A$ is the fraction of contact that is refreshed. Refreshing occurs when tilts change local normal stresses, thereby removing aged contacts and adding new ones. New contacts do not contain information about the loading history, and thus the memory of the interface is partially erased. Complete erasure corresponds to $\phi = 1$, and the resulting memory-less interface would evolve as a single logarithm, as is the case after the interface is slid. Consistent with the model, we find $\phi \propto |\Delta S |$, as shown in Fig \ref{3}(c). However, the calculated magnitude of $\phi$ is too high to stem only from refreshing of contacts due to a change in local normal stress. A change in shear on the order of half of the normal load ($\Delta S / N = 1/2$) produces $\phi \sim 0.4$. According to our model this would indicate that almost half of the contact area has relocated, which is clearly not the case, as was shown in Fig \ref{2}(a) for $\Delta S/N = 4/9$.

The unadulterated effect of a change in shear can be measured only by examining the central region of an interface, which experiences almost no change in normal stress when the shear load is modified. We find that in this region, $\phi \propto |\Delta S | $ as well, as shown by the black symbols in Fig \ref{3}(c). That is to say, even without torque-induced tilts, a change in shear still erases a significant fraction of the local memory of the interface. One appealing potential physical mechanism for this erasure is a slip occurring for only a portion of the interface \cite{Berthoud:1999ha}. Two factors discredit this possibility. First, we see no visual evidence of local slip. Second, this is inconsistent with the fact that shear does not affect the frictional aging rate when tilts are minimized.  Nevertheless, this behavior remains in line with the phenomenological description of the interface as consisting of an ensemble of exponential modes of relaxation with a broad distribution of timescales \cite{Amir:2012fe, Lahini:2017dq, Dillavou:2018in}. In this picture, a change in shear would allow the system to access new modes of relaxation while retaining the gains of some old modes. This may be possible through irreversible (plastic) deformation of contacts, locking-in previously reversible deformations and generating a new micro-structure and thus a new set of relaxation modes.

A striking and perhaps useful corollary of shear-induced access to new modes is confirmed experimentally. When tilts are minimized, a rapid shift in shear does not reduce total contact but temporarily accelerates the growth of $A_R$, as shown in Fig \ref{3}(b)-(c). Therefore while any constant $S_0$ gives the same aging rate in both $\beta^A$ and $\beta^\mu$, these rates can be boosted through a change in shear load. This temporary boost can be made continuous by periodically cycling between two shear loads the system, which increases both $\beta^A$ and $\beta^\mu$ as a result, as shown in Fig \ref{3}(d). 

We have shown that shear-enhanced aging is a consequence of minute interfacial tilts and erasure of local memory. The growth rate of the real area of contact under static normal load, $\beta^A$, is insensitive to the presence of a static shear load. However, a \textit{change} in shear load redistributes interfacial contact by inducing miniscule relative tilts between the two surfaces. As a result, $\beta^\mu$ is linearly dependent on $S_0$, extending into negative shear load (pulling). Minimizing these tilts eliminates the effect of a static shear load on $\beta^\mu$ entirely. Finally, we have shown that tilts are not the whole story; even under no-tilt conditions, a change in shear partially erases interfacial memory. This erasure is non-destructive, and can be harnessed to increase the growth of $\mu_S$ and $A_R$. 

The shear-accelerated aging in frictional interfaces may be analogous to aging in other amorphous materials such as disordered networks, disordered holey sheets, and polydisperse packing of disks \cite{Pashine:2019wl}. In these systems, aging has a directional component, even though the systems are isotropic. For example, the Poisson's ratio in a foam network measured by compressing along a given axis is markedly different if the system is aged under compression along that same axis versus along a perpendicular axis. In our system, when the shear load is changed, the direction of the local stress vector at every contact changes. Thus, the system may be accessing different modes of relaxation, just like in foam networks and granular materials, when compressed along different axes. Repeatedly changing the shear gives the system access to more modes of relaxation, and aging accelerates as a result.

\begin{acknowledgments}
\end{acknowledgments}
\bibliography{bib}

\end{document}